\documentclass[12pt]{article}

\usepackage{graphicx}  
\usepackage{multicol}  
\usepackage{epsfig}   
\usepackage{subfigure} 

\newcommand{\fig}[1]{Fig.~\ref{#1}} 

\newcommand{\tabl}[1]{Table~\ref{#1}} 
\newcommand{\eqn}[1]{Eq.~(\ref{#1})}
\newcommand{\secn}[1]{Section~\ref{#1}}

\begin{document}
\thispagestyle{empty}

\vspace{1cm}
\begin{center}

{\Large{\bf Deleting species from model food webs}}

\vspace{1cm}

{\it Christopher Quince$^{1}$, Paul G. Higgs$^2$ and Alan J.
McKane$^{3}$}

\vspace{1cm}

$^1$Department of Physics and Astronomy, Arizona State University, \\
Tempe, AZ 85287-1504, USA \\

$^2$Department of Physics and Astronomy, McMaster University, \\ 
Hamilton ON, Canada L8S 4M1

$^3$Department of Theoretical Physics, University of Manchester, \\
Manchester M13 9PL, UK \\

\end{center}

\begin{abstract}
We use food webs generated by a model to investigate the effects of deleting
species on other species in the web and on the web as a whole. The model 
incorporates a realistic population dynamics, adaptive foragers and other 
features which allow for the construction of model webs which resemble 
empirical food webs. A large number of simulations were carried out to 
produce a substantial number of model webs on which deletion experiments 
could be performed. We deleted each species in four hundred distinct model 
webs and determined, on average, how many species were eliminated from the 
web as a result. Typically only a small number of species became extinct; 
in no instance was the web close to collapse. Next, we examined how the 
the probability of extinction of a species depended on its relationship with 
the deleted species. This involved the exploration of the concept of
indirect predator and prey species and the extent that the probability of 
extinction depended on the trophic level of the two species. The effect of 
deletions on the web itself was studied by searching for keystone species, 
whose removal caused a major restructuring of the community, and also by 
looking at the correlation between a number of food web properties (number 
of species, linkage density, fraction of omnivores, degree of cycling and 
redundancy) and the stability of the web to deletions. With the exception 
of redundancy, we found little or no correlation. In particular, we found 
no evidence that complexity in terms of increased species number or links 
per species is destabilising. 
\end{abstract}

\newpage

\section{Introduction}

The world's ecosystems are increasingly being subjected to stresses that
result in large-scale changes in species population densities. These stresses
often directly or indirectly arise from human activities and include
pollution, over-exploitation, species invasions and habitat destruction 
(Carlton and Geller 1993, Milner-Gulland and Bennett 2003). Understanding how
ecosystems respond to such perturbations is therefore highly important. 

Here we will attempt to contribute to this understanding by focusing on 
species deletion, the complete removal of a species from an ecosystem 
community, using the theoretical framework of dynamical modelling. Species 
deletion is a large-scale perturbation of particular relevance, as it is
a commonly used empirical tool to measure interactions strengths within real
communities, and can be considered a reasonable approximation to
other large perturbations (Paine 1980, Pimm 1980). 

The theory of small perturbations in dynamical models of ecosystems is well 
developed. It began with May's seminal work showing that the probability 
of an ecosystem with random interactions being locally stable decreases 
with both the number of species, the frequency of interactions and the 
strength of those interactions (May 1972, 1973). This result was an 
important contribution to the complexity-stability debate and contradicted 
earlier ideas that complexity should naturally lead to stability (Odum 1953, 
MacArthur 1955, Elton 1958). A crucial conceptual element to May's work 
is that only a local knowledge of the dynamics, encapsulated in the 
``community matrix'', is necessary to determine the stability of the 
population equilibrium to small perturbations. This is also true of 
perturbations that actually alter the position of the equilibrium, provided 
that they are small enough (Yodzis 1989). The results of small perturbations 
can be determined using only ``local models'', where the species growth 
rates are approximated by linear functions of the population densities 
(Yodzis 2001).

In contrast the study of large-scale perturbations requires a ``global
model'' --- one defined over the whole of phase space. The use of such a 
model will inevitably involve a modelling choice, but it is important that 
it incorporates phenomena such as non-linear functional responses and 
adaptive foraging, which will be likely to operate over these 
large changes in population density (Abrams 1996). 

A global model with these features has already been developed as part of a 
larger model of community coevolution by some of us (Drossel et al. 2001). 
This model assembles ecosystem communities through the repeated 
addition of new species that are modified versions of those already present. 
It is therefore conceptually similar to community assembly models (Drake 1990, 
Law and Morton 1996, Morton and Law 1996, Lockwood et al. 1997), the 
difference being that new species are generated {\it in situ} rather than
being taken from a species pool. In the model species are constantly 
being subjected to large perturbations in population densities as new 
species add and existing species go extinct. Crucially, species are 
allowed to alter their foraging strategies in response to these changes.
Thus the population dynamics of the model are particularly well suited to the 
study of species deletion and will be used in this study. Since previous 
studies of deletion have used the Lotka-Volterra equations or the equivalent 
discrete time Ricker dynamics, this will give a unique perspective on the 
problem (Pimm 1979, Pimm 1980, Borrvall et al. 2000, Lundberg et al. 2000).

We will not only use the population dynamics of the model, we will also use 
it to generate the food webs from which species will be deleted. It might 
be preferable to use real food web structures, but then interaction strengths 
that possess a stable equilibrium for that structure would have to be 
determined, which is likely to be difficult for a large food web. The only 
studies of deletion that have used large empirical food webs circumvent 
this problem by adopting a network approach, without using an explicit 
population dynamics (Sol\'e and Montoya 2001, Dunne et al. 2002). Their 
procedure reveals interesting features of the network topology, but by using 
the simple rule that species go extinct when none of their prey remain, they 
ignore the complex realities of the dynamics. Our model allows us to 
generate a data set of food webs with reasonably realistic topologies 
and interaction strengths (Drossel et al. 2001, Quince et al. unpublished),
 suitable for investigating the complex dynamics of deletion in 
multi-species communities.

The outline of the rest of the paper is as follows. We begin with a review 
of the model in \secn{model}, followed by the details of the generation 
of the food web data set and the deletion experiments themselves in 
\secn{data}. The next sections concern the results of these experiments.
In  \secn{trophicrelationships} we investigate how the trophic relationship
between a pair of species influences the outcome of deleting one of them, 
in \secn{keystonespecies} we go on to consider which species properties 
correlate with large changes in community composition following that species 
removal and in \secn{stability} we consider the stability of the communities 
as a whole to deletion and the extent to which this is determined by their 
food web structure. We finish with a brief discussion of the major 
results in \secn{conclusion}.

\section{The model}
\label{model}

We will now give a short description of the model we will use to evolve the 
food webs to be used in our study. Further details are given in Caldarelli 
et al. (1998), Drossel et al. (2001), Quince et al. (2002) and Quince et al.
 (unpublished).

The dynamics of the model has a different appearance depending on the 
time-scales under consideration. On time-scales of the order of the lifetimes 
of individuals, the number of species is fixed and the dynamics is 
deterministic: it is given by a variant of the standard equations of 
population dynamics. This includes the capacity of all species to change their 
predator-prey characteristics with time, competition between predators and a
varying amount of effort that a predator puts into catching a particular prey. 
The system is allowed to change according to this dynamics for what may be 
relatively long intervals until equilibrium is reached, that is, until the 
populations of the different species present remains unchanged. If during 
the population dynamics the density of a species falls below a value 
$N^{\rm min}$, usually taken to be 1, it is assumed to have become extinct, 
and is removed from the system. 

Once an equilibrium of the populations densities is reached a speciation 
event is initiated: a new species is generated by changing 
one of the features of one of the individuals of a randomly chosen species.
This new species is then added to the system with a population density 
$N^{\rm child}$, also taken to be 1 in these simulations. The system is then 
again allowed to develop under the deterministic equations of the population 
dynamics. The small population of the new species may give rise to a 
viable population, or it may die out, but eventually when a new 
equilibrium is reached, a new speciation event will take place. By repeating 
this procedure tens of thousands of times an entire food web can be evolved, 
using a combination of conventional population dynamics and stochastic 
speciation events. On these very long evolutionary time-scales the discrete 
time steps where speciation occurs are the main aspect of the dynamics. 

In order to define the speciation process, and more generally to be able to
characterise a species, we need to introduce features which when taken 
together make up a species. In our model, features are specified by integers:
$\alpha = 1,\ldots, K$. Any subset of $L$ of these features constitutes a
species. It is assumed that the effectiveness of predator-prey relationships 
among species is due to the effectiveness of individual features against 
each other. Therefore the score of one species, $i$, against another, $j$, 
denoted by $S_{ij}$, is defined in terms of the $K \times K$ matrix 
$m_{\alpha \beta}$ which gives the score of feature $\alpha$ against feature
$\beta$:
\begin{equation}
S_{ij} = {\rm max}\left\{ 0, \frac{1}{L}\,\sum_{\alpha \in i}\,
\sum_{\beta \in j} m_{\alpha \beta} \right\}\,.
\label{score}
\end{equation}
The matrix $m_{\alpha \beta}$ is antisymmetric. Its independent elements are 
random Gaussian variables with zero mean and unit variance chosen at the 
beginning of a simulation run and not changed during that particular run. 
This allows the score of one species against another to be calculated from 
(\ref{score}): if $S_{ij} > 0$ then species $i$ is adapted for predation 
against species $j$, if $S_{ij}=0$ then it is not. We will also need to
define the overlap $q_{ij}$, between two species $i$ and $j$,
as the fraction of features of species $i$ that are also possessed by 
species $j$. The external environment is represented by a species indexed 0.
This is assigned a random set of $L$ features at the 
beginning of a run, and is not changed throughout the course of the run. 

Having described the structure of a species in the model, and used this to 
define the score (\ref{score}) and overlap, we
will now use these quantities in the construction of the population dynamics
that governs the changes in population sizes between speciation events. The 
rate of change of $N_{i}(t)$, the population size of species $i$ at time $t$, 
is given by
\begin{equation}
\frac{dN_{i}}{dt} = - N_{i} + \lambda \sum_{j} N_{i} g_{ij} - \sum_{j}
N_{j} g_{ji} \,.
\label{popdyn}
\end{equation}
The function $g_{ij}$ is the functional response: the rate at which one 
individual of species $i$ consumes individuals of species $j$. The choice 
of $g_{ij}$ essentially defines the nature of the population dynamics. We
will give the explicit form chosen below, but it is here that the dependence 
on the score and the overlap functions will enter. It will also depend on
the $N_{k}(t)$, and so will change with time. 

The terms on the right-hand side of (\ref{popdyn}) are now simply 
interpreted. The last factor represents the loss in resources for species 
$i$ due to predation by all of the other species (in the model the measure of 
resources and species number are synonymous). The factor 
$\sum_{j} N_{i} g_{ij}$ on the other hand represents the gain to species $i$
from predation on the set of species $j$ including the environment, species 0.
The environment is assigned a fixed population, $N_{0} = R/\lambda$, thus 
$R$ is a parameter of the model that controls the rate of input of external 
resources. If it is assumed that a fraction $\lambda$ of 
the resources gained through predation are used to create new members of 
species $i$, the second term on the right-hand side of (\ref{popdyn}) is 
obtained. Finally, the first term simply represents the rate of death of 
individuals in the absence of interaction with other species. 

In order to briefly motivate the form of the functional response we will use,
let us first discuss the case of a single predator $i$ feeding on a single 
prey $j$. In this case 
\begin{equation}
g_{ij}(t) = \frac{S_{ij} N_{j}(t)}{b N_{j}(t) + S_{ij} N_{i}(t)}\,,
\label{simple}
\end{equation}
where $b$ is a constant. We can gain more understanding of the structure of
$g_{ij}$ by noting that when the predators are far more numerous than the 
prey ($N_{i} \gg N_{j}$), $N_{j} \sim g_{ij}N_{i}$: the feeding rate of the 
predators is limited only by the number of prey. In the other limit, when the
prey is very abundant compared with the predators ($N_{j} \gg N_{i}$),
$g_{ij} \sim S_{ij}/b$: each predator feeds at a constant maximum rate. This
latter result also gives an interpretation to the constant $b$. Having 
introduced the basic form (\ref{simple}), we can now state general form for 
the functional response used in the model:
\begin{equation}
g_{ij}(t) = \frac{S_{ij} f_{ij}(t) N_{j}(t)}
{b N_{j}(t) + \sum_{k} \alpha_{ki} S_{kj} f_{kj}(t) N_{k}(t)}\,.
\label{gij}
\end{equation}
There are two new aspects present in (\ref{gij}) and absent in (\ref{simple}):
\begin{itemize}
\item[1.]Interference competition between predators of prey species $j$ is 
modelled by the factor $\alpha_{ki}$. We take $\alpha_{ii}=1$ and 
$\alpha_{ki} < 1, i \neq k$ to reflect the fact that competition between 
members of the same species is typically stronger than competition between 
different species. In fact, we expect that the more species are alike, the
greater will be the competition between them, and therefore take
\begin{equation}
\alpha_{ij} = c + (1 - c)q_{ij}\,,
\label{alpha}
\end{equation}
where $c$ is a constant lying between zero and one which is the residual 
degree of competition that exists even if two competing predators have no 
features in common.

\item[2.] Adaptive foraging is modelled using the factors $f_{ij}$.
The effort $f_{ij}$ can be viewed as the fraction of time an individual of 
species $i$ spends predating on species $j$ or the fraction of the population 
of species $i$ dedicated to consuming only $j$. These efforts must satisfy 
$\sum_{j} f_{ij} = 1$ for all $i$. To determine the $f_{ij}$ it seems 
reasonable to assume that the gain which an individual of species $i$ makes 
in consuming individuals of species $j$ (that is, $g_{ij}$), divided by the 
amount of effort $i$ puts into this task (that is, $f_{ij}$), should be the 
same for all prey species $j$. Using this condition, together with the 
normalisation of the $f_{ij}$, leads to 
\begin{equation}
f_{ij}(t) = \frac{g_{ij}(t)}{\sum_{k} g_{ik}(t)}\,.
\label{ESS}
\end{equation}
This choice of efforts can be shown to be an evolutionary stable strategy 
(Drossel et al. 2000), or in the terminology of foraging theory, an 
ideal free distribution of predators across prey (Fretwell and Lucas 1970).

\end{itemize} 

The calculation of the efforts, through (\ref{ESS}), effectively introduces 
a new behavioural timescale into the problem. We assume that the efforts 
change on a much shorter timescale than the population densities $N_{j}(t)$, 
and therefore that they may be found by iterating (\ref{gij}) and (\ref{ESS}) 
assuming constant population densities. When this process has been completed, 
we may then move on to updating the population densities. Thus we do not 
treat the efforts as dynamical variables, instead we assume that they are a 
function of the population densities, even if we have no explicit form for 
this function. This describes the essential features of the model. In the 
rest of the paper we will explore the consequences of performing deletion 
experiments on food webs created by the model.

\section{Deletion experiments}
\label{data}


\begin{figure}[t]
\begin{center}
\includegraphics[width=6.5cm]{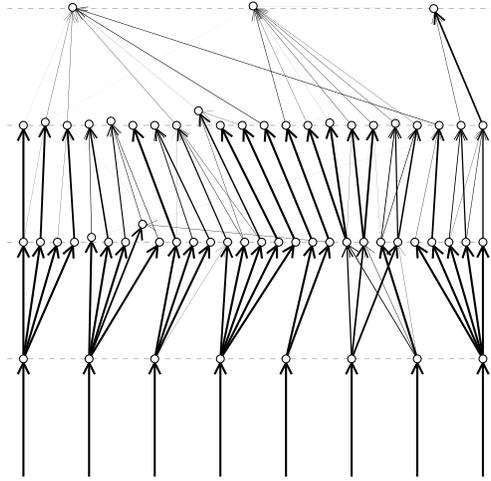}
\end{center}
\caption[Typical food web]{A typical model food web.}
\label{typical}
\end{figure}


The set of model food webs used in this study was obtained by performing
four hundred independent simulations of the model, each simulation
lasting for 120000 speciations. These simulations were independent, in that
different pseudo-random number sequences were used in their generation. Thus
the simulations differed in their random matrices, $m_{\alpha \beta}$, 
environment features and speciation events. The same parameters were used 
in all the simulations these being $R = 1 \times 10^5$, $b = 5\times10^{-2}$, 
$c = 0.5$, $\lambda = 0.1$ and $N^{\rm min} = N^{\rm child} = 1.0$. The 
effect of altering the model parameters is investigated in Quince et al. 
(unpublished). The four hundred final food webs from these simulations, which after 
120000 speciations will have structures drawn from a stationary distribution, 
constituted the ecosystem data set.

A typical food web is shown in \fig{typical}. Each species in this diagram 
is represented as a circle, the sizes of which are the same for all species.
This differs to the convention adopted in Quince et al. (2002), where 
the radii of the circles were proportional to the logarithm of the population 
densities. The arrows represent predator-prey interactions, with the arrow 
pointing from the prey to the predator. The intensity of the arrow is 
proportional to the fraction of the predators diet that consists of that 
particular prey. The vertical arrows originating from the base of the 
diagram, rather than from another species, indicate that the species is 
feeding off the environment. The species are positioned vertically according 
to trophic height, defined as the average path length from the species to 
environment, the average being weighted by predator diet fractions. The 
dashed lines show the position of integer values of the trophic height.


\begin{table}[t]
\begin{center}
\begin{tabular}{|c|c|c|c|} \hline
Statistic & Symbol & Mean & Std. Dev. \\
\hline \hline 
Number of species & $S$ & 63.8275 & 7.5152\\
\hline
Links per species & $L/S$ & 1.6881 & 0.1258\\
\hline
Fraction of omnivores& $O$ & 0.1606 & 0.0474\\
\hline
Degree of cycling& $\mathcal{C}$ & 0.0049 & 0.0022\\
\hline
Redundancy & $\mathcal{r}$ & 0.1884 & 0.0706\\
\hline \hline
\end{tabular}
\end{center}
\caption[]{The means and standard deviations of five food web statistics 
for the four hundred model food webs in the data set. \label{stats}}
\end{table}


The four hundred food webs in the model data set span a range of structures. 
We shall quantify this variation with five food web statistics: 
\begin{enumerate}
\item  The total number of species in the food web denoted by $S$. 
\item The links per species denoted by $L/S$. This quantity is simply the 
number of predator-prey interactions divided by the number of species, where 
we will use the convention of counting a link if it constitutes greater 
than 1\% of a predator's diet. 
\item The fraction of omnivorous species in the food web denoted by $O$. We 
define omnivorous species as those which feed at more than one trophic level 
and define the trophic level of a species to be the shortest path from that 
species to the environment. The justification for this choice is the 
observation (Yodzis 1984) that the shortest path between a species and the 
environment tends to be the most important energetically. 
\item The degree of cycling denoted by $\mathcal{C}$. The method we adopt 
for measuring the amount of cycling of energy in the food webs is based on 
both the ideas presented in Ulanowicz (1983) and on source code kindly 
provided by the author. Essentially a backtracking algorithm was first used 
to identify each cycle in a food web. Having done this, the amount of energy 
flowing in a cycle was identified with the strength of the weakest link of 
the cycle, exactly as in Ulanowicz (1983). To define $\mathcal{C}$ we measure 
the proportion of a predator's prey obtained through cyclic flow averaged 
over the species in the web.
\item The ecosystem redundancy denoted by $\mathcal{r}$. The redundancy of 
an ecosystem is the proportion of species which can be considered 
superfluous to the functioning of the ecosystem as a whole (Walker 1992). We 
define a species to be redundant in our food webs if at least one other 
species possesses the same pattern of trophic links i.e. the same predators 
and prey. As for the calculation of $L/S$, only links forming greater than 
1\% of the predator's diet are used in this calculation. Then the redundancy, 
$\mathcal{r}$, is the fraction of redundant species in the food web.
\end{enumerate}
In \tabl{stats} the means and standard deviations of the five statistics 
are shown for the four hundred food webs in the data set.
 
\begin{figure}[t]
\begin{center}
\rotatebox{0}{\scalebox{2.5}{\includegraphics[width=.3\textwidth]
{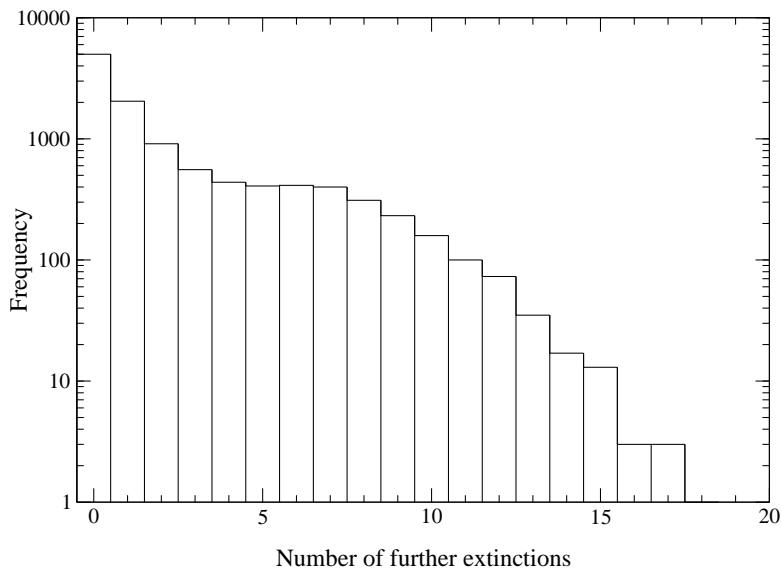}}}
\end{center}
\caption[Distribution of further extinction number]{The frequency distribution of the number of further extinctions 
for the 25531 species deletions.} 
\label{FEDist}
\end{figure}

This data set was then used to investigate the effect of species deletion.
A single species was removed from a web and the population dynamics iterated 
until a stable equilibrium was reached. If, during this process, the 
population of a species fell below $N^{\rm min}$, then it was removed (``went 
extinct'') in accordance with the criterion applied when evolving the 
communities. For each web every species was deleted independently, that is, 
the webs were returned to their original state between deletions. In total, 
over all the food webs, 25531 species were deleted. In \fig{FEDist} we show 
the frequency distribution of number of further extinctions for these 
deletions. This distribution does not decay exactly exponentially with 
further extinction number, but it does have a characteristic size of just a
few species. The number of further extinctions seems bounded, the largest 
number is seventeen, and on no occasion is the whole web of typically sixty 
species close to collapse. The effect of deleting a species is localised in 
the webs.


\section{Trophic relationships and species deletion}
\label{trophicrelationships}

In this section we investigate how the trophic relationship between 
a pair of species influences the impact that deleting one of the pair will 
have on the other. We began by simply taking the four hundred webs described 
in the previous section and placing each species into four non-exclusive 
categories according to its trophic relationship with the deleted species. 
These categories were predators, prey, competitors and indirect predators. 
The definitions of the first two categories are obvious. A ``competitor'' 
was defined as any species that shared a prey with the deleted species. 
The final category of ``indirect predators'' requires a little more 
explanation. Consider any two species in an ecosystem, $i$ and $j$. If 
there exists one or more paths from $j$ to $i$ travelling only up trophic 
links, from prey to predator, and the minimum length of those paths 
is greater than one, then we define species $i$ to be an indirect predator 
of species $j$. In other words, species $i$ is an indirect predator of $j$ 
if some of the resources consumed by $i$ came originally from $j$, but species 
$i$ does not prey on $j$ directly. In calculating the categories all links 
which formed less than one percent of the predators diet were ignored.

Having defined these categories we calculated two quantities. The first was 
the mean number of species in each category that went extinct following 
species deletion, with the average performed over every deleted species in 
every web. The second was the probability that a species with a particular 
trophic relationship to the deleted species went extinct. This quantity 
was estimated by simply averaging the proportion of species with a 
particular relationship that went extinct. These results are shown in 
\tabl{trophicr}. The first row shows the results for all species, in order to 
aid comparison with the other results. We see that, on average, following 
the deletion of a species from a web, a further 1.366 species went extinct 
and that any given species had a probability of 0.021 of going extinct. A 
clearer picture is usually obtained by using probabilities, rather than 
mean values, when making comparisons between tropic relationships. This is 
due to the weaker dependence that probabilities have on the number of 
species with that particular relationship to the deleted species. From 
these results we see that that a predator of a deleted species is more than 
ten times as likely to go extinct following the deletion of its prey than 
an average species in the web. This makes obvious intuitive sense. More 
interestingly it is also found that a prey of the deleted species is twice 
as likely to go extinct as an average species in the web when its predator 
is removed. The most likely explanation for this is that an effect known as 
predator-mediated coexistence or keystone predation is operating. This is 
a mechanism whereby a predator allows inferior competitors to coexist with 
a superior competitor by predation of the competitively dominant species. 
Thus the deletion of the predator can lead to extinctions amongst its prey. 
This has been observed in real communities (Paine 1974, Lubchenco 1978, 
Navarrete and Menge 1996) and in theoretical studies of simple dynamical 
systems (Fujii 1977, Shigesada and Kawasaki 1988).

The plausibility of the above mechanism is supported by the fact that 
competitors of the deleted species are five times less likely to go 
extinct than an average species, suggesting that competition plays an 
important part in structuring these communities. The effect on indirect 
predators is also significant: they are five times as likely to go 
extinct following the removal of their indirect prey compared to the average. 
It is interesting to note that the effect on an indirect predator is less than
that on a direct predator. We will return to this later.      


\begin{table}
\begin{center}
\begin{tabular}{|c|c|c|}
\hline Trophic relationship&Mean number of 
extinctions&Probability of extinction\\ 
\hline \hline
All species&1.366&0.021\\
\hline  
Predators&0.675&0.233\\
\hline
Prey&0.132&0.041\\ 
\hline
Competitors&0.013&0.004\\ 
\hline
Indirect predators&0.658&0.106\\
\hline \hline
\end{tabular}
\end{center}
\caption{The effect of trophic relationship on the probability and mean 
number of species going extinct following deletion of a species from an 
ecosystem.}
\label{trophicr}
\end{table}


\subsection{Predator-prey interaction strengths and extinction probabilities}

The results in \tabl{trophicr} show that removing the prey, or 
indirect prey, of a species has a negative effect on the predator. Clearly 
the importance of a prey species to its predator will vary between 
predator-prey pairs. A measure of the positive effect that a prey species has 
on the predator will be the effort $f_{ij}$, corresponding to the fraction
of the prey in the predators diet (Ulanowicz and Puccia 1990).

We can derive a similar quantity for the positive effect that an indirect 
prey species has on its indirect predator. Consider the square of the $f$ 
matrix $f_{ij}^2=\sum_{k=0}^Nf_{ik}f_{kj}$, where the sum is over all species 
in the web and $k=0$ corresponds to the environment. Since $f_{kj}$ is the 
fraction of species $k$'s diet that comes from species $j$, then $f_{ik}f_{kj}$
is the fraction of species $i$'s diet that comes from species $j$ via species 
$k$. If we now sum $k$ over all species in the web, we obtain the fraction of 
species $i$'s diet that comes from species $j$ through all paths of length 2. 
Therefore if we define the matrix $F = \sum_{n=1}^{\infty} f^n$, then 
$F_{ij}$ is the fraction of species $i$'s diet that comes from species $j$ 
via all possible paths. Since all resources originally derive from the 
environment, $F_{i0} = 1$ for any species $i$. For the direct matrix 
$f_{ij}$, normalisation ensures that $\sum_{j=0}^Nf_{ij} = 1.0$. However for 
the indirect matrix,
\begin{equation}
\sum_{j=0}^{N} F_{ij} = \sum_{j=0}^{N} \left\{ f_{ij} + \sum_{n=2}^{\infty}
(f^{n})_{ij} \right\} \geq \sum_{j=0}^{N} f_{ij} = 1.0 ,
\label{Fij}
\end{equation}
with equality only when species $i$ is a basal species.

If the infinite series of matrices converges, it can be calculated using 
$F = (1 - f)^{-1}f$. All eighty $f$ matrices in this study were such that the
sum was finite, however we have not been able to prove the convergence of the
sum for a general web.

We are interested in separating the effects of indirect and direct predation. 
This can be achieved by defining a further matrix $I$ such that 
\begin{equation}
I_{ij} =
\left\{ \begin{array}{ll}
F_{ij}, & \mbox{\ if $f_{ij}=0.0$} \\
\ 0 \ , & \mbox{\ if $f_{ij}>0.0$\, .}
\end{array} \right.
\label{I_ij}
\end{equation}


Thus if $I_{ij}>0$, then species $i$ will be an indirect predator of species 
$j$ according to the definition given above.

\begin{figure}[t]
\begin{center}
\rotatebox{0}{\scalebox{2.5}{\includegraphics[width=.3\textwidth]
{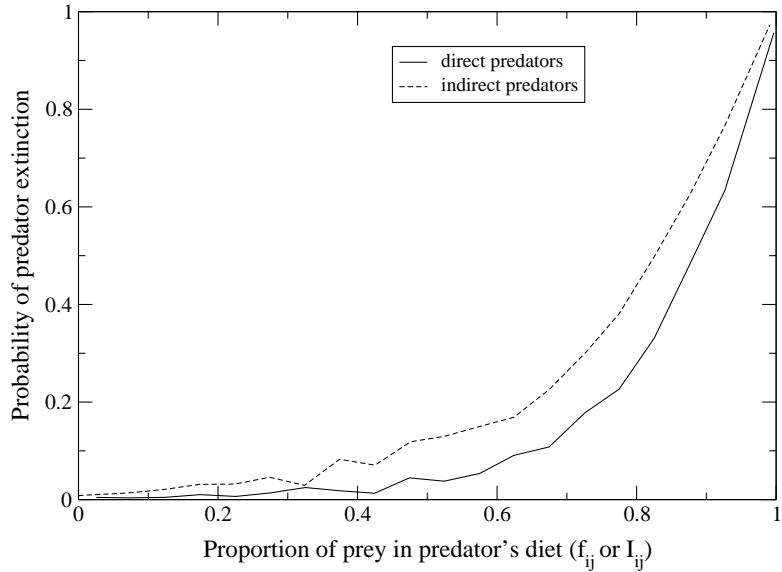}}}
\end{center}
\caption[The effect of diet contribution on predator extinction]{The 
fraction of predators (indirect predators) going extinct upon prey removal 
as a function of the proportion of that prey (indirect prey) in the predators 
diet.} 
\label{PredExtinct}
\end{figure}


We can now examine how the probability that a predator or indirect predator 
goes extinct, following removal of its prey, varies with the proportion 
of the prey in the predators diet. This was carried out by dividing all the 
predator-prey (indirect predator-prey) pairs in the four hundred ecosystems 
into bins of size 0.05 according to their associated $f_{ij}$ ($I_{ij}$) 
values. The proportion of prey removals that resulted in the extinction
of the predator (indirect predator) was then calculated for each bin and 
plotted in \fig{PredExtinct}.

The interpretation of \fig{PredExtinct} is quite straightforward. 
Ignoring fluctuations attributable to sampling effects, both curves are 
monotonic. As the fraction of the predators, or indirect predators, diet 
that is obtained from the deleted species increases, so does the 
probability of predator extinction. However the probability of extinction 
does not become large until the prey constitutes a significant fraction of 
the predators diet. For direct predators it does not reach 10\% until 
$f_{ij} \approx 0.65$ and for indirect predators this occurs when 
$I_{ij} \approx 0.45$. This probably arises from incorporating adaptive 
foraging into the population dynamics: predators can survive events that 
remove a large portion of their prey. In both cases the extinction 
probability rapidly approaches, but does not quite reach, 100\% as 
$f_{ij}$ ($I_{ij}$) approaches 1.0. In fact, when a prey constitutes greater 
than 99\% of a predators diet its removal leads to predator extinction only 
97\% of the time. This again illustrates the effect of adaptive foraging: 
3\% of the time the predator must be surviving by exploiting a prey species 
that previously formed less than 1\% of its diet.


\begin{figure}[t]
\begin{center}
\rotatebox{0}{\scalebox{2.5}{\includegraphics[width=.3\textwidth]
{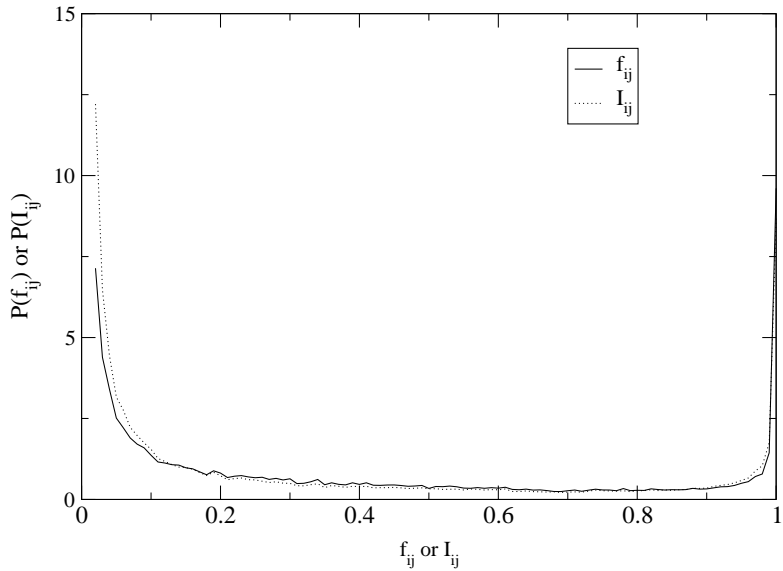}}}
\end{center}
\caption[]{The probability distribution of $f_{ij}$ and $I_{ij}$ for all 
$f_{ij} > 0.01$ and $I_{ij} > 0.01$. These results are a compilation 
over the four hundred food webs described in \secn{data}.}
\label{distfI}
\end{figure}


The two curves in \fig{PredExtinct} have a similar form, but at any given
$f_{ij}$ or $I_{ij}$ value the probability of extinction is greater 
for indirect predators. This seems to contradict 
\tabl{trophicr}, which shows a greater probability of deletion for 
direct predators than indirect predators. However this can be explained by 
the distribution of the non-zero elements of the $I$ and $f$ matrices. If 
there are more small values of $I_{ij}$ than $f_{ij}$, then defining a 
predator (indirect predator) as having $f_{ij} > 0.01$ ($I_{ij} > 0.01$) will 
lead to a lower probability of deletion for indirect predators than direct 
predators, even though for a given value of $f_{ij}$ ($I_{ij}$) the effect of 
removing the indirect predator is more significant. That this is the case 
can be seen from \fig{distfI}, where the probability distributions of 
$f_{ij}$ and $I_{ij}$ for all $f_{ij} > 0.01$ and $I_{ij} > 0.01$ are shown.
This does not of course explain why an indirect predator is more likely to go
extinct when prey constituting a given fraction of its diet is removed.
This may arise from the assumption that the net importance of multiple
paths to the predator can be obtained by simply summing their individual 
weights implicit in \eqn{Fij}.


\begin{figure}[t]
\begin{center}
\rotatebox{0}{\scalebox{2.5}{\includegraphics[width=.3\textwidth]
{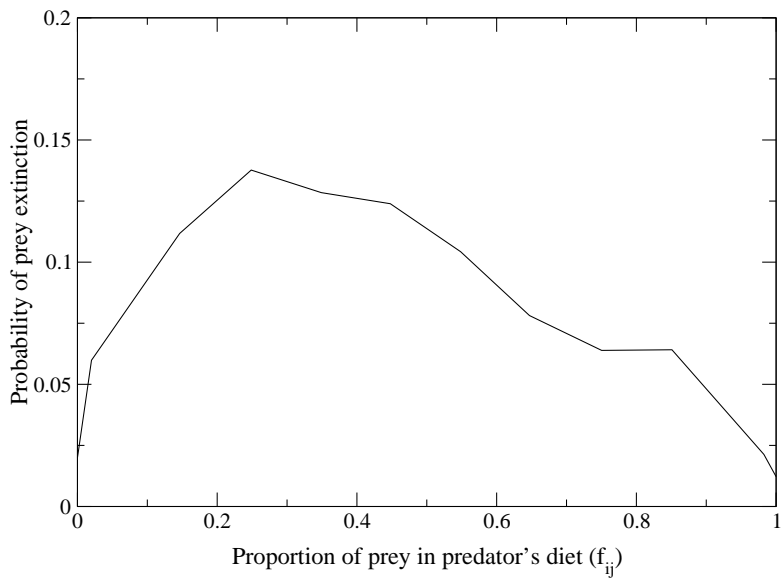}}}
\end{center}
\caption[The effect of diet contribution on prey extinction]{The fraction of 
prey going extinct upon predator removal as a function 
of the proportion of that prey in the predators diet.} 
\label{PreyExtinct}
\end{figure}


We showed in \tabl{trophicr} that a prey species has an increased 
probability of extinction following removal of its predator. We might also 
expect that this effect will depend on the fraction of the prey in the 
predators diet. This is investigated in \fig{PreyExtinct}, where the 
probability of prey extinction following predator removal has been estimated 
as a function of $f_{ij}$, by placing all the predator-prey interactions 
from the four hundred food webs into bins of size 0.1, and calculating the 
proportion of deletions that led to extinction of the prey. From this 
graph we see that prey extinction probability peaks at an intermediate 
value around $f_{ij} = 0.25$, where slightly less than 15\% of prey are 
going extinct. This is what we would expect if the mechanism for prey 
extinction is predator-mediated coexistence as proposed above, since if a 
prey forms only a fraction of a predators diet, then that predator is also
likely to be exploiting its competitors. 


\begin{table}
\begin{center}
\renewcommand{\arraystretch}{1.4}
\setlength\tabcolsep{3pt}
\begin{tabular}{|c||c|c|c|c|c|}
\hline Level& All & 1 & 2 & 3 & 4\\ 
\hline
\hline All&1.366&0.011&0.649&0.679&0.026\\
\hline 1&6.571&0.012&3.651&2.783&0.124\\
\hline 2&0.705&0.013&0.121&0.560&0.011\\
\hline 3&0.544&0.009&0.359&0.161&0.015\\
\hline 4&1.223&0.006&0.498&0.709&0.010\\
\hline
\end{tabular}
\end{center}
\caption{The mean number of species on a given level that go extinct as a 
function of the trophic level of the deleted species. The rows refer to the 
trophic level of the deleted species and the columns to the trophic level 
in which further extinctions occurred.}
\label{levels}
\end{table}


\subsection{Trophic Levels}

In \tabl{trophicr} we categorised species according to their trophic 
relationship to the deleted species. Another way to categorise species in 
a food web is by trophic level. As mentioned in \secn{data} we use the 
minimum path length definition of trophic level (Yodzis 1984). 
Having assigned a trophic level to each species in our ecosystems, we 
calculated the mean number of species on each level that went extinct as a 
function of the trophic level of the deleted species. These results are 
shown in \tabl{levels}, where the same averaging procedure was used as 
for \tabl{trophicr}, that is, the average was performed over all the 
deleted species in all eighty ecosystems. 

The analysis of this data is aided by statistics on how the number of 
species $S$, the average prey number $S_{\rm prey}$, and average predator 
number $S_{\rm predator}$, vary between trophic levels. The number of 
species $S$ in each trophic level was calculated by averaging over the 
eighty webs in this study. The other statistics were averaged over the 
total number of species in each trophic level. These statistics together 
with standard deviations in brackets are shown in \tabl{levelstat}.


\begin{table}
\begin{center}
\renewcommand{\arraystretch}{1.4}
\setlength\tabcolsep{5pt}
\begin{tabular}{|c||c|c|c|}
\hline Level&$S$&$S_{\rm prey}$& $S_{\rm predator}$\\ 
\hline  
\hline 1&7.76(1.06)&0(0)&4.26(1.34)\\
\hline 2&30.19(4.41)&1.13(0.37)&1.68(0.74)\\
\hline 3&24.63(4.27)&2.71(1.62)&1.25(0.49)\\
\hline 4&1.26(0.84)&5.84(3.31)&1.09(0.29)\\ 
\hline \hline
\end{tabular}
\end{center}
\caption{The dependence of number of species $S$, number of prey 
$S_{\rm prey}$, and number of predators $S_{\rm predator}$, on trophic 
level. The figures in brackets are standard deviations.}
\label{levelstat}
\end{table}


Several patterns emerge from \tabl{levels}. For example, if we remove 
a species from a particular level then this leads to extinctions on the level 
above and this effect diminishes as we increase the level of the 
deleted species. Thus the mean number of level 2 species going extinct upon 
removal of a level 1 species is 3.651, which is roughly seven times the mean 
number of level 3 species going extinct following a deletion on level 2 
(0.560), which in turn is about forty times larger than the mean number of 
extinctions in level 4 as a result of a deletion on level 3 (0.015). We can 
explain this with reference to \tabl{levelstat}, from which we see that 
while the mean number of prey increases rapidly with trophic 
level, the mean number of predators decreases. These two effects 
will be complementary, so that as we increase the trophic level of the deleted 
species, the number of predators of that species decreases, and those 
predators become less specialised on the deleted species. This leads to 
less extinctions in the level above when the species is removed. As a 
concrete example, consider deletion of a level 1 species. This will, on 
average, affect about 4 species on level 2 and these 4 species will be almost 
entirely dependent on the deleted species (as mentioned previously, deletion 
of the only prey of predator leads to extinction of that predator almost 
100\% of time). Thus we would expect about 4 predators in level 2 to go 
extinct as a result of a deletion on level 1. This compares well with the 
actual value of 3.651. We also see from \tabl{levels} that the effect 
of deleting a species on the level above propagates up trophic levels, so 
that removing a level 1 species leads to deletions on level 2, which in turn 
leads to extinction on levels 3 and 4. This is as we would expect, given that
we have already illustrated that consequences of species deletion can propagate
up food webs. 

As well as extinctions occurring in the levels above a deleted 
species, we find that deleting a species can lead to extinctions in the 
level below it. The magnitude of this effect seems to decrease with the 
trophic level of the deleted species: the mean number of level 3 species 
going extinct given deletion on level 4 is 0.709, compared to 0.359 for the 
number of level 2 species going extinct as the result of a deletion in 
level 3. Similarly, the deletion of a level 2 species leads to an average of 
just 0.013 extinctions on level 1. These observations can be explained if
we assume that predator-mediated coexistence causes the extinctions, since
the more prey a predator has, the stronger we might expect this effect to be.
Then the pattern of increasing extinctions in the level below, with increasing
level of the deleted species, merely reflects the increase in average prey 
number with trophic level observed in \fig{levelstat}. The question of why
the model food webs show these changes in mean predator and prey number 
with trophic level is not addressed here, but is discussed in
 Quince et al. (unpublished).

\section{Keystone species}
\label{keystonespecies}

Thus far the focus of this study has been on pairs of species and how 
the trophic relationship between them influences the effect that deleting 
one species will have on the other. It was found that the effects of 
deletion can propagate both up and down food webs. We now consider the 
related question of what factors determine the impact that deleting a 
species will have on the whole web. Specifically, we ask whether there are any 
consistent differences between species whose removal causes little change 
in the food web, and those which play a major role in structuring the 
community. The latter are sometimes referred to as ``keystone species'', 
although this term can be restricted to those species whose importance is 
large relative to their population size (Power et al. 1996). Here we 
will simply be interested in identifying factors that statistically 
influence the number of further extinctions that follow the removal of 
a species.

The trophic level of the deleted species has an effect on the expected 
number of further extinctions. This is shown in the first column of 
\tabl{levels}, where the mean number of further extinctions in the web as 
a whole is seen to be largest for species on trophic level 1. This probably 
reflects the greater importance of effects propagating up, rather than down,
the food web. An alternative way to categorise species according to trophic 
role, which is better at separating these two processes, is into the 
following three classes: `top' (species with no predators), `intermediate' 
(species with both predators and prey) and `basal' (species with no prey). 
When this is done, it is found that removal of a basal species causes an 
average of 6.58 further extinctions as opposed 0.65 and 0.61 for intermediate 
and top species respectively. Comparing the figures for basal and top species
reveals that top-down processes are indeed statistically less important 
than bottom-up in these communities.


\begin{figure}[t]
\begin{center}
\rotatebox{0}{\scalebox{2.5}{\includegraphics[width=.3\textwidth]
{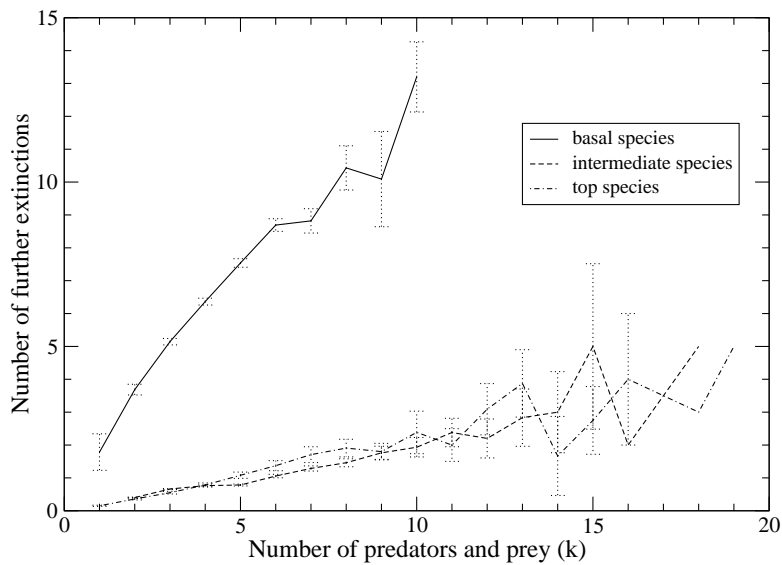}}}
\end{center}
\caption[The relationship between link number and expected number of 
further extinctions]{The number of further extinctions following the deletion of a species 
as a function of its total number of predators 
and prey also known as the node degree $k$. The results are subdivided 
according to whether the deleted species is basal, intermediate or top 
and are averaged over all species in all four hundred communities. The 
error bars give standard errors in the mean.
} 
\label{KD}
\end{figure}


The above categories are quite broad. One property which we can use to more 
finely discriminate between species is the number of other species they 
interact with through predator-prey links. We might expect this to 
correlate with the impact of deleting the species on the food web. That 
this is indeed the case is shown in \fig{KD}, where the average number 
of further extinctions is plotted as a function of the node degree $k$ in 
network terminology. In the case of food webs the latter corresponds to the 
number of predators plus the number of a prey of a species. The results 
are shown for basal, intermediate and top species separately. For all 
three classes the average number of further extinctions increases with $k$, 
so the most connected species are the ones whose removal has the greatest 
effect on the food web structure. The results for top species provide 
further support, albeit circumstantial, that the top-down effect in our webs
is predator-mediated coexistence, since we would expect the importance of 
this effect to increase with the number of prey of the predator. They can 
be contrasted with an earlier study that failed to find such a 
relationship (Pimm 1980).

These results have relevance for the studies of real food web robustness 
to deletion mentioned in the introduction. These obviated the need for 
an explicit dynamics by considering multiple removals, and judging a further 
extinction to have occurred when all the prey of a species are absent
(Sol\'e and Montoya 2001, Dunne et al. 2002). These studies found that 
removing the most connected species resulted in more secondary extinctions 
for the same number of species removed. Our results suggest that if a 
population dynamics was included in these studies, the food web structures 
would be even more sensitive to the removal of highly connected species. 


\begin{figure}[t]
\begin{center}
\rotatebox{0}{\scalebox{2.5}{\includegraphics[width=.3\textwidth]
{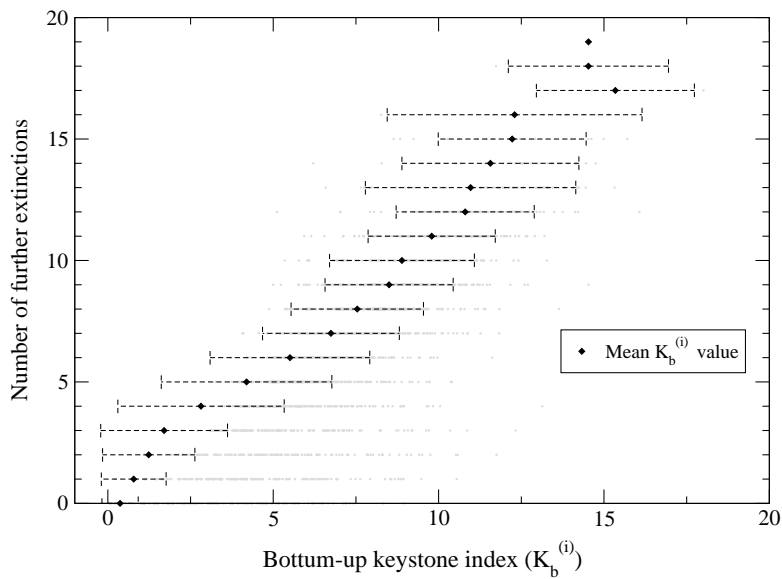}}}
\end{center}
\caption[The relationship between bottom-up keystone index and expected 
number of further extinctions]{The number of further extinctions following the 
deletion of a basal species $i$ plotted against its bottom-up keystone 
index ($K_{b}^{i}$). The faint grey dots give individual data points 
present in order to give a sense of the distribution. The black diamonds give 
means with error bars showing standard deviations of the distribution of 
$K_{b}^{i}$ values for each number of further extinctions.} 
\label{KbDBasal}
\end{figure}



\begin{figure}[t]
\begin{center}
\rotatebox{0}{\scalebox{2.5}{\includegraphics[width=.3\textwidth]
{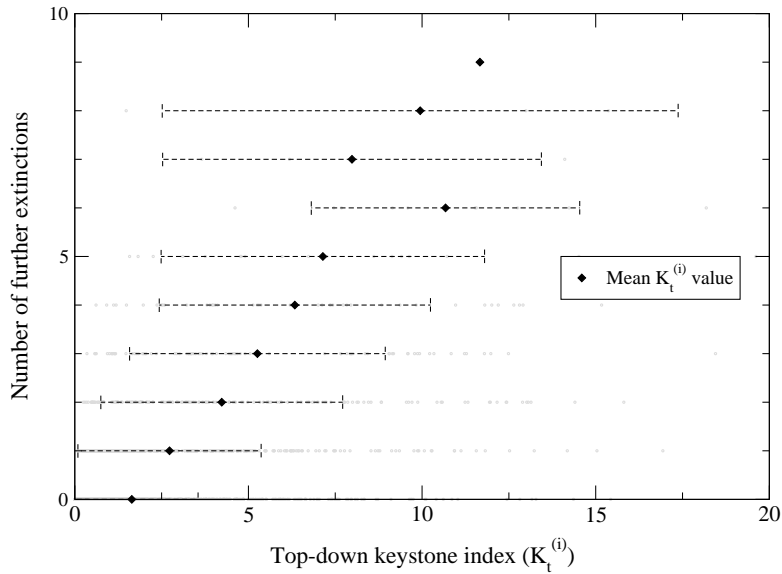}}}
\end{center}
\caption[The relationship between bottom-up keystone index and expected 
number of further extinctions]{The number of further extinctions following the 
deletion of a top species $i$ plotted against its top-down keystone index 
($K_{t}^{i}$). The faint grey dots give individual data points and the black 
diamonds give means with error bars showing standard deviations of the 
distribution of $K_{t}^{i}$ values for each number of further extinctions.} 
\label{KtDTop}
\end{figure}


The correlation between $k$ and further extinction number is very strong, 
especially given that it quantifies the number of direct interactions of 
a species, and does not give information on the strength of indirect effects, 
which we have already shown to be important (Jordan and Scheuring 2002). 
It is difficult to devise measures of species importance that do incorporate 
indirect effects. Potential candidates are the bottom-up and top-down 
keystone indices of Jordan et al. (1999). These were originally 
devised for binary food webs, but is easy to extend them to the model food webs
including diet compositions considered here. In fact the bottom-up keystone 
index for a species $i$ is simply
\begin{equation}
K_{b}^{i} = \sum_{j = 1}^{S} F_{ji}\,.
\end{equation}
Here the matrix $F$ is defined by $F = \sum_{n=1}^{\infty} f^n$ and, as 
discussed in the previous section, its elements give the fraction of species 
$i$'s diet that comes from species $j$ via all possible paths. This makes 
the meaning of $K_{b}^{i}$ clear: it is the total number of species that 
depend on $i$ directly or indirectly for resources. The complementary 
quantity $K_{t}^{i}$ measures the strength of top-down effects it can be 
defined in a similar way for non-binary food webs, which we will not discuss 
here. There is in fact a very good correlation between $K_{b}^{i}$ and 
further extinction number for basal species. This shown in \fig{KbDBasal}.
This suggests that this quantity is effective at predicting the strength of 
bottom-up effects. We also found a correlation between $K_{t}^{i}$ and the 
number of further extinctions following the removal of a top species 
(\fig{KtDTop}) although not as strong as for basal species. Furthermore the 
composite quantity $K^{i} = K_{t}^{i} + K_{b}^{i}$ did have some success at 
predicting the outcome of removing an intermediate species. This suggests 
that although top-down effects might be more difficult to quantify with 
these network measures, and they are not yet as effective as simply counting 
the number of predators and prey, they may well prove a practical way of 
predicting the importance of species in food webs.

\section{Stability of food webs to deletion}
\label{stability}

\begin{figure}[t]
\begin{center}
\includegraphics[width=6.5cm]{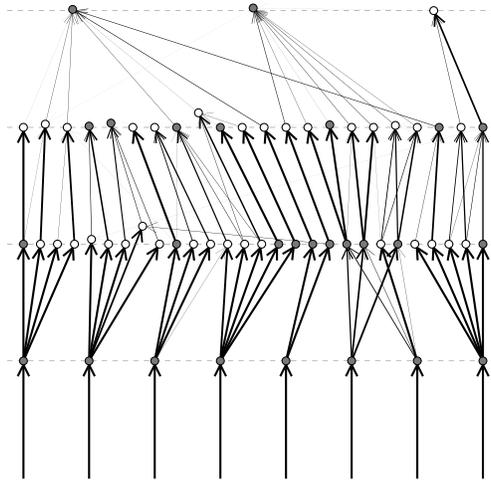}
\end{center}
\caption{The food web of \fig{typical} with those species whose deletion 
causes further extinctions shaded grey. There are 27 such species. The 
number of species in the web is 61, giving a deletion stability, 
$S_{d} = 34/61$ or 0.557. The latter is a fairly typical value for the 
webs in this study.}
\label{deletion}
\end{figure}

In the previous section we focused on the effect that deleting one species 
has on another in the model food webs. Here we will consider the stability of 
the communities as a whole. In particular we will ask whether that 
stability is determined by the topological structure of the ecosystem. We 
begin by defining deletion stability as the fraction of species that have 
the property that, when they alone are deleted, further extinctions do not 
take place. This definition is similar to that given by Pimm (1979), and 
has the advantage that it should not have any in-built dependence on 
species number. Its calculation is illustrated in \fig{deletion}. 


We will use the four hundred model food webs detailed in \secn{data} as 
our data set and describe their topological structure with the five food 
web statistics also defined there. These five statistics quantify four 
food web properties that it has been suggested may impact ecosystem 
stability: complexity in terms of number of species and linkage density 
($S$, $L/S$), the amount of omnivory in the web ($O$), the importance 
of cycles ($\mathcal{C}$) and the redundancy ($\mathcal{r}$). We remind 
the reader that the means and standard deviations of these statistics for 
the data set are given in \tabl{stats}. 

If we possessed a range of food webs that vary for each property 
independently, whilst the other properties remain fixed, then this analysis 
would be quite simple. Indeed this was the approach adopted in previous 
studies using small pre-defined food web structures (Pimm 1979, Pimm 1980,
 Borrvall et al. 2000). However by generating the food webs 
by evolving a number of communities at a particular set of parameter values, 
we are faced with a less straightforward situation. The food webs vary for 
all properties simultaneously, and some of the properties are significantly 
correlated with one another. In fact, the situation is much like analysing 
real food web structures. The advantage is that our range of structures 
are much more complex and realistic, the disadvantage is that we are forced 
to adopt a statistical approach.


\begin{table}[t]
\begin{center}
\renewcommand{\arraystretch}{1.4}
\setlength\tabcolsep{5pt}
\begin{tabular}{|c|c|c|c|c|c|c|}
\hline
& $S_{d}$ & $S$ & $L/S$ & $O$ & $\cal C$ & $\cal r$\\
\hline \hline
$S_{d}$ & --- & -0.054 & 0.063 & -0.162* & -0.111 & 0.197* \\
\hline
$S$ & --- & --- & 0.354* & 0.073 & -0.129 & 0.034  \\
\hline
$L/S$ & --- & --- & --- & 0.197* & -0.196* &  -0.264*  \\
\hline
$O$ & --- & --- & --- & --- & 0.359* &-0.282*\\
\hline
$\cal C$ & --- & --- & --- & --- & --- & -0.085 \\
\hline\hline
\end{tabular}
\end{center}
\caption{The correlation matrix for the variables: deletion stability 
($S_{d}$), number of species ($S$), links per species ($L/S$), fraction 
of omnivores ($O$), degree of cycling ($\mathcal{C}$) and redundancy 
($\mathcal{r}$). The values given are the linear correlation coefficients 
$r$ between the pairs of variables calculated over the 400 food webs. The 
correlations judged to be significant, those with a probability $p$ of 
no correlation smaller than the Bonferroni corrected value of 0.05/15, are 
highlighted with an asterisk.}
\label{correlationmatrix}
\end{table}


We will commence our analysis by examining  the correlation matrix for all 
the variables, both the dependent variable $S_{d}$, and food web properties. 
This is shown in \tabl{correlationmatrix}. The deletion stability is 
significantly correlated with the fraction of omnivores (negatively) and 
the redundancy (positively). However because of the many significant 
correlations between the variables themselves, we can not conclude that a 
smaller proportion of omnivores or greater redundancy will be associated 
with higher deletion stability, all other properties being unchanged. 


\begin{table}
\begin{center}
\begin{tabular}{|c|c|c|c|c|}
\hline
Variable & $\rm b_{j} $ & Std. err. & $t$ & $p$\\
\hline\hline
Intercept** & 0.4157 & 0.0803 & 5.1766 & 0.0000\\
\hline
$S$*       & -0.0016 & 0.0007 & -2.2903 & 0.0225\\
\hline
$L/S$**     & 0.1436 & 0.0453 & 3.1738 & 0.0016\\
\hline
$O$*       & -0.2512 & 0.1183 & -2.1231 & 0.0344\\
\hline
$\cal C$  & -1.4098 & 2.4820 & -0.5680 & 0.5703\\
\hline
$\cal r$**  & 0.3043 & 0.0749 & 4.0634 & 0.0001\\
\hline \hline
\end{tabular}
\end{center}
\caption{The multivariate linear regression of deletion stability ($S_{d}$) as 
a function of number of species ($S$), links per species ($L/S$), fraction of 
omnivores ($O$), degree of cycling ($\cal C$) and redundancy ($\cal r$). This 
fit had ${\cal R}^2 = 0.08186$, ${\cal F} = 7.026$ on 5 and 394 degrees of 
freedom, and $P = 2.636\times10^{-06}$. The coefficients that are 
significantly different from zero ($p < 0.05$) are marked with an asterisk, 
and those that are highly significant ($p < 0.01$) with a double asterisk.
The calculation was performed using the software package S-Plus 6.0. (Mathsoft Inc. 2000)}
\label{multipleregression}
\end{table}


One way to shed some light on this problem is to use multivariate linear 
regression (Jobson 1991). This statistical procedure essentially assumes 
that the dependent variable, in this case $S_{d}$, can be described as a 
linear function of the independent variables, the five food web properties 
plus a noise term. The results of such an analysis are shown in 
\tabl{multipleregression}. Examining the regression coefficients, $\rm b_{j}$,
for the individual variables we see that two, the redundancy again and the 
links per species, are deemed to have a strong influence on the deletion 
stability, and that both of these effects are positive. More pertinently 
however, examining the fit as whole we see from the ${\cal R}^2$ value that 
only 8\% of the variation in stability is being explained by these variables.

We therefore conclude that for this data set a fairly robust positive 
relationship exists between the proportion of redundant species in the web 
and the stability of the web to deletion. This result makes sense. We defined
a redundant species as one that is functionally equivalent, in the sense of 
possessing the same predators and prey, to at least one other species. Thus 
the removal of a redundant species is unlikely to cause further species to 
go extinct, since its functional equivalents should be able to increase 
their population sizes and compensate for the loss. This supports the 
hypothesis that increased redundancy in ecosystems will result in increased 
functional reliability (Walker 1992, Naeem 1998). 

We find no evidence that complexity in terms of increased species number 
or links per species is destabilising. In fact there was evidence that 
increasing the latter actually reduced the probability of further extinctions 
when species are deleted. This can be compared to early work on deletion 
stability in small food webs modelled with Lotka-Volterra dynamics, where it 
was found that increasing either the number of species or connectance, 
$L/S^{2}$, rapidly decreased stability (Pimm 1980). The difference can 
probably be attributed to the more realistic structures and global dynamics 
used here, in particular incorporating adaptive foraging into the population 
dynamics. 

This result can also be compared to more recent work examining the effect 
of deleting species from simple three level food webs, constructed such that 
all species on a trophic level were functionally equivalent (Borrvall
 et al. 2000). These authors found that stability to deletion increased 
with the number of species on each trophic level. Their set-up corresponds to 
keeping $\mathcal{r}$ constant and equal to one, whilst increasing the total 
number of species, $S$, a variable for which no effect was observed here. 
This study also used Lotka-Volterra dynamics and this may explain the
discrepancy. In particular, they observed that removing a predator had no 
effect on its prey, in contrast to the results detailed in 
\secn{trophicrelationships}.

We also found that omnivory and cycling were unimportant in determining 
deletion stability. In fact, all our structural properties taken together 
explained very little of the variation between the webs. It may be that 
the measure of deletion stability we used is intrinsically noisy --- it does 
seem sensitive to the presence of one or two vulnerable species --- or it 
may be that our statistics are not capturing the properties that are 
important in determining robustness to deletion. 

\section{Conclusion}
\label{conclusion}

In this paper we have shown that deletion experiments, which are very
difficult and time-consuming to carry out in real communities, can be 
easily implemented on model webs, which previous studies have shown have 
many of the characteristics of real webs (Drossel et al. 2001).
We found a number of interesting results on species removal from communities 
which differed from some previous studies. The food webs as a whole were 
shown to be quite robust to deletions, and individual species were able to 
survive the loss of prey constituting a major fraction of their diet. 
Deletions were shown to cause further extinctions amongst species both above 
and below the deleted species in the food web. These phenomena arose out of 
the complex population dynamics used in the model which incorporates adaptive 
foraging. They illustrate the importance of using a realistic global dynamics 
when considering community responses to large scale perturbations such as 
deletion. This contrasts with studies that either lack an explicit dynamics 
(Sol\'e and Montoya 2001, Dunne et al. 2000) or use Lotka-Volterra 
equations (Pimm 1979, Pimm 1980, Borrvall et al. 2000, Lundberg et al. 2000). 

In addition to a realistic dynamics, effective studies of species removal 
require a realistic set of structures. By using a set of large webs composed 
of species spanning a range of trophic roles, we were able to show that 
removing the most connected species resulted in the most further extinctions 
and that recently developed `keystone indices' were fairly effective at 
predicting species importance. A study using small simple webs failed
to find these relationships (Jordan et al. 2002). The range of food 
web structures studied allowed us to show the important role that redundant 
species play in increasing food web robustness to deletion, and find that 
there was no correlation between increased complexity and decreased stability 
to deletion. This adds another component to the stability-complexity debate.

This paper represents the first attempt to join complex food web structures 
with realistic population dynamics to study species loss from communities.
It would be interesting to see if the results would change if real, rather 
than evolved, food web structures were used with such a dynamics or if the
particular choice of dynamics were changed. In any case, we believe that 
we have shown that it is crucial for models to display a degree of realism,
if reliable deductions concerning the consequences of species deletions are 
to be made.

\vspace{0.9cm}

\noindent{\bf Acknowledgements}: We would like to thank R. Ulanowicz for 
letting us have the source code for his program on cycling in food webs. 
We also wish to thank R. Law and M. Rattray for useful discussions. CQ thanks
the EPSRC (UK) for financial support during the initial stages of this work.

\section*{References}

\noindent Abrams, P. A. 1996. Dynamics and interactions in food webs 
with adaptive foragers. - In: Polis, G. A. and Winemiller, K. O. (eds),  
Food webs: integration of patterns and dynamics. Chapman \& Hall, pp.\,113-121.

\noindent Borrvall, C., Ebenman, B. and Jonsson, T. 2000. Biodiversity lessens 
the risk of cascading extinction in model food webs. - Ecol. Lett. 3: 131-136.

\noindent Caldarelli, G., Higgs, P. G. and McKane, A. J. 1998. Modelling 
coevolution in multispecies communities. - J. Theor. Biol. 193: 345-358.

\noindent Carlton, J. T. and Geller, J. B. 1993. Ecological roulette -- the 
global transport of nonindigenous marine organisms. - Science 261: 78-82.

\noindent Drake, J. A. 1990. The mechanics of community assembly and 
succession. - J. Theor. Biol. 147: 213-233.

\noindent Drossel, B., Higgs, P. G. and McKane, A. J. 2001. The influence of 
predator-prey population dynamics on the long-term evolution of food web 
structure. - J. Theor. Biol. 208: 91-107.

\noindent Dunne, J. A., Williams, R. J. and Martinez, N. D. 2002. Network 
structure and biodiversity loss in food webs: robustness increases with 
connectance. - Ecol. Lett. 5: 558-567. 

\noindent Elton, C. S. 1958. The ecology of invasions by animals and plants.
 - Methuen, London.

\noindent Fretwell, D. S. and Lucas, H. L. 1970. On territorial behaviour 
and other factors influencing habitat distribution in birds. - Acta 
Biotheor. 19: 16-32.

\noindent Fujii, K. 1977. Complexity-stability relationship of 
two-prey-one-predator species system model: local and global stability.
 - J. Theor. Biol. 69: 613-623.

\noindent Jobson, J. D. 1991. Applied multivariate data analysis.
 - Springer-Verlag.

\noindent Jordan, F., Takacs-Santa, A. and Molnar, I. 1999. A reliability 
theoretical quest for keystones. - Oikos 86: 453-462.

\noindent Jordan, F. and Scheuring, I. 2002. Searching for keystones in 
ecological networks. - Oikos 99: 607-612.

\noindent Jordan, F., Scheuring, I. and Vida, G. 2002. Species positions 
and extinction dynamics in simple food webs. - J. Theor. Biol. 215: 441-448.

\noindent Law, R. and Morton, R. D. 1996. Permanence and the assembly of 
ecological communities. - Ecology 77: 762-775.

\noindent Lockwood, J. L., Powell, R. D., Nott, P. and Pimm, S. L. 1997. 
Assembling ecological communities in time and space. - Oikos 80: 549-553.

\noindent Lubchenco, J. 1978. Plant species-diversity in a marine inter-tidal 
community: importance of herbivore food preference and algal competitive 
abilities. - Am. Nat. 112: 23-39.

\noindent Lundberg, P., Ranta, E. and Kaitala, V. 2000. Species loss leads to 
community closure. - Ecol. Lett. 3: 465-468.

\noindent MacArthur, R. H. 1955. Fluctuations of animal populations and a 
measure of community stability. - Ecology 36: 533-536.

\noindent Mathsoft Inc. 2000. S-Plus software: version 6.0. - MathSoft Inc., Cambridge, MA. 

\noindent May, R. M. 1972. Will a large complex system be stable?
 - Nature 238: 413-414.

\noindent May, R. M. 1973. Stability and complexity in model ecosystems. 
 - Princeton Univ. Press.

\noindent Milner-Gulland, E. J and Bennett, E. L. 2003. Wild meat: the bigger 
picture. - Trends Ecol. Evol. 18: 351-357. 

\noindent Morton, R. D. and Law, R. 1997. Regional species pools and the 
assembly of local ecological communities. - J. Theor. Biol. 187: 321-331.

\noindent Navarrete, S. A. and Menge, B. A. 1996. Keystone predation and 
interaction strength: interactive effects of predators on their main prey. 
 - Ecol. Monogr. 66: 409-429.

\noindent Naeem, S. 1998. Species redundancy and ecosystem reliability. 
 - Conserv. Biol. 12: 39-45.

\noindent Odum, E. P. 1953. Fundamentals of ecology. - Saunders.

\noindent Paine, R. T. 1974. Intertidal community structure: experimental 
studies on the relationship between a dominant competitor and its principal 
predator. - Oecologia 15: 93-120.

\noindent Paine, R. T. 1980. Food webs : linkage, interaction strength and 
community infrastructure. - J. Anim. Ecol. 49: 667-685.

\noindent Pimm, S. L. 1979. Complexity and stability -- another look at 
MacArthur's original hypothesis. - Oikos 33: 351-357.

\noindent Pimm, S. L. 1980. Food web design and the effect of species 
deletion. - Oikos 35: 139-149.

\noindent Power, M. E., Tilman, D., Estes, J. A., {\it et al.} 1996. 
Challenges in the quest for keystones. - Bioscience 46: 609-620.

\noindent Quince, C., Higgs, P. G. and McKane, A. J. 2002. Food web structure 
and the evolution of ecological communities. - In: L\"assig, M. and 
Valleriani, A. (eds), Biological Evolution and Statistical Physics. 
Springer-Verlag, pp.\,281-298.

\noindent Shigesada, N. and Kawasaki, K. 1988. Direct and indirect effects of 
invasions of predators on a multiple-species community. - Theor. Pop. Biol. 
36: 311-328.

\noindent Sol\'e, R. V. and Montoya, J. M. 2001. Complexity and fragility in 
ecological networks. - Proc. Roy. Soc. Lond. B268: 2039-2045.

\noindent Ulanowicz, R. E. 1983. Identifying the structure of cycling in 
ecosystems. - Math. Biosci. 65: 219-237.

\noindent Ulanowicz, R. E. and Puccia, C. J. 1990. Mixed trophic impacts in 
ecosystems. - Coenoses 5: 7-16.

\noindent Walker, B. H. 1992. Biodiversity and ecological redundancy.
 - Conserv. Biol. 6: 18-23.

\noindent Yodzis, P. 1989. Introduction to Theoretical Ecology.
- Harper \& Row. 

\noindent Yodzis, P. 2001. Must top predators be culled for the sake of 
fisheres? - Trends Ecol. Evol. 16: 78-84.

\end{document}